\documentclass[twocolumn,letterpaper]{IEEEAerospaceCLS}  % only supports two-column, letterpaper format
\usepackage[T1]{fontenc}
\usepackage{graphicx}
\usepackage[colorlinks=true, urlcolor=blue!60!black, linkcolor=blue!60!black, citecolor=blue!60!black]{hyperref}
\usepackage[numbers, square]{natbib}
\usepackage{placeins}
\usepackage{subcaption}
\usepackage{amsmath}
\usepackage{amssymb}
\usepackage{booktabs}
\usepackage{array}
\usepackage{multirow}
\usepackage{algorithm}
\usepackage{algpseudocode}
\usepackage{xcolor}
\usepackage{siunitx}
\usepackage{xcolor}
\usepackage{listings}
\usepackage[capitalise]{cleveref}
\crefname{lstlisting}{Listing}{Listings}
\crefname{appendix}{Appendix}{Appendices}
\let\linkedciteauthor\citeauthor
\renewcommand{\citeauthor}[1]{{\hypersetup{citecolor=black}\linkedciteauthor{#1}}}
\begin{document}
\title{Where Planning Changes the Outcome of Orbital Games}

\author{%
Duncan Eddy\\
Stanford University\\
496 Lomita Mall\\
Stanford, CA 94305\\
deddy@stanford.edu
\and
Mahdi Al-Husseini\\
Stanford University\\
496 Lomita Mall\\
Stanford, CA 94305\\
mah9@stanford.edu
\and
Mykel J. Kochenderfer\\
Stanford University\\
496 Lomita Mall\\
Stanford, CA 94305\\
mykel@stanford.edu
%\thanks{\footnotesize 979-8-3195-1048-8/27/$\$31.00$ \copyright2027 IEEE}
}

\maketitle

\thispagestyle{plain}
\pagestyle{plain}

\begin{abstract}
This paper introduces OrbitalGym, a simulation environment for studying adversarial orbital encounters to enable understanding of when planning algorithms or system design determine game outcomes. We consider the problem of planning maneuvers for rendezvous and proximity operations scenarios where two non-cooperative teams of satellites compete to complete a differential-reward objective, referred to as ``orbital game''. Recent advances in terrestrial decision making for game playing have led to an influx of new opportunities and proposed methods for solving orbital games. However, studies typically evaluate one method in one scenario configuration, often with a passive or random defender. It is unclear whether improved planning method matters once vehicle capability (maneuverability, sensing, communications) and challenging adversarial responses are considered.

We create a JAX-based environment for modeling orbital games that accommodates varying numbers of agents, state representations (position and velocity, attitude, and fuel), communication mechanisms (an explicit action or an implicit, state-dependent ability), and sensor models (proximity-based or conical field-of-view). OrbitalGym compiles games to batched GPU accelerator code, enabling large-scale, parallel simulation of control policies across wide ranges of configurations and initial conditions. This enables identification of regimes where policy choice can have a significant effect on outcomes versus regimes where system capability dominates the result, regardless of policy choice.

We use OrbitalGym to study where the choice of policy changes the outcome of two close-range games: pursuit-evasion, where one spacecraft attempts to capture another, and lady-bandit-guard, where a guard prevents an attacker from reaching a protected asset. We compare different planning methods including classical PD control, optimal LQR control, model predictive path integral control, and Monte Carlo tree search over different ratios of thrust capabilities, propellant budgets, initial separations, game duration, sensor capabilities, and the definition of capture. In pursuit-evasion the ratio of propellant budgets decides most encounters: against five of six evader policies, the choice of pursuer policy changes the capture probability by 10\% or more in at most 15 of 90 capability configurations. In lady-bandit-guard, the choice of policy has a much larger effect: against bandits capable of reaching the protected asset, a better guard policy increases defense probability by at least 10\% in 56–74 of 90 capability configurations. All code is available at \url{https://github.com/sisl/OrbitalGym}.
% Hack to fit table to contents on same column as abstract
\vspace{-3.8em}
\end{abstract}

{\hypersetup{linkcolor=black}\tableofcontents}

\section{Introduction}\label{sec:introduction}
Rendezvous and proximity operations (RPO) are becoming more common in commercial and government missions. Servicing, inspection, and debris-removal satellites maneuver to within close proximity of other spacecraft, however the same maneuvers can serve an adversarial purpose. Between January 2024 and March 2025, there were more than 6 instances of Chinese and Russian satellites maneuvering within close proximity of other nations' satellites without warning or explanation~\cite{swope2025}. An observer who sees such a maneuver cannot easily tell whether it is benign or if a response is required. Conflict in space is costly for every party. Ultimatum bargaining theory, from the study of international conflict, finds that when conflict is costly and both sides are fully informed about each other, the probability of conflict goes to zero. Uncertainty about the opponent is what makes conflict likely~\cite{reed2003}. This motivates a transparent study of adversarial decision-making in orbit: by thorough study of the problem, including how system differences and planning decisions affect the outcome of a given scenario, every operator can be fully informed of the result in advance, and obviate the need for conflict. Despite this, the problem remains understudied. Existing work typically solves single instances: one scenario, system configuration, and decision algorithm, often against a passive or scripted opponent. However, there is a large configuration space for orbital games that might affect outcomes, spanning vehicle capability (maneuverability, sensing, communications), encounter geometry and duration, and the decision methods of both sides. The configuration space is too large to understand one instance at a time. Studying it requires an environment in which many configurations, scenarios, and algorithms can be evaluated in the same way at low cost.

The space community has studied adversarial encounters between spacecraft as differential games since the first decade of spaceflight. \citeauthor{wong1967} solved planar intercept games in a constant gravity field~\cite{wong1967}. Later work found saddle-point solutions of the three-dimensional orbital pursuit-evasion game with numerical methods~\cite{pontani2009,shen2018}, solved minimax trajectories in the Hill-Clohessy-Wiltshire (HCW) frame~\cite{stupik2012}, used sensitivity methods to locate barrier surfaces that separate capture from escape~\cite{hafer2015}, and bounded capture by the reachable sets of two-impulse transfers at large separation~\cite{venigalla2021}. Feedback guidance laws avoid learning a new policy for every encounter type~\cite{menon1988,jagat2017,kim2025}. Recent studies bring the methods of the planning community to these games: applying tree search over impulsive maneuvers~\cite{xie2024}, deep reinforcement learning for interception~\cite{zhang2022} or multi-agent pursuit~\cite{zhao2023,wang2024}. SpaceGym~\cite{allen2023} and AstroCraft~\cite{anderson2026} provide training and evaluation environments for select orbital games.

Recently, the planning and learning community has significantly advanced decision-making for complex, high-dimensional state-space games when paired with scalable environments built to simulate them. Monte Carlo tree search (MCTS) made sampling-based lookahead practical in large games~\cite{kocsis2006uct,browne2012mcts}. AlphaGo combined tree search with learned policy and value networks to surpass grandmaster play in Go~\cite{silver2016}, AlphaZero learned the same games from self-play alone~\cite{silver2018alphazero}, MuZero extended these methods with a learned model~\cite{schrittwieser2020muzero}, and AlphaStar reached professional-level play in StarCraft II~\cite{vinyals2019alphastar}. Each result relied on a simulation environment that allowed millions of trial games. Gymnasium, and PettingZoo standardized the interfaces for policy training and evaluation for single-agent and multi-agent environments and enable easy deployment of new algorithms across different games~\cite{brockman2016gym,terry2021pettingzoo}.

The two lines of work are complementary. Methods from the planning community transfer to orbital games, and modeling decisions for a specific scenario, including its associated satellite systems, determine what those methods can achieve. Each study above assumes a configuration---an initial geometry, a set of maneuvering capabilities,  a sensor model (or perfect information), a game duration, and an opponent decision-policy, and reports how well a method performs within it. Several orbital game studies vary a few of these parameters~\cite{xie2024,zhao2023,anderson2026}, however all assume perfect information shared by all actors, which creates uncertainty about whether the algorithmic improvements persist under more realistic assumptions. For instance, a better guidance algorithm changes nothing when the sensor loses the target. A comparison of decision methods for orbital games therefore should state its configuration precisely and test whether its performance holds under different conditions. The question this paper asks is where the choice of policy changes the outcome of an orbital game, and where capability, information, or the definition of the game decides it.

We introduce OrbitalGym\footnote{Source code is available under an MIT License and can be found at \url{https://github.com/sisl/OrbitalGym}.}, a simulation environment for five different orbital games, and use it to answer this question for two games at close range: pursuit-evasion (PE), in which one spacecraft tries to capture another, and lady-bandit-guard (LBG)~\cite{rusnak2005}, in which a guard blocks an approaching bandit from reaching a protected asset. We next compare performance of classical feedback laws, including proportional-derivative (PD) control and the linear-quadratic regulator (LQR), a model predictive path integral (MPPI) planner~\cite{williams2017mppi}, on both sides of each game, and MCTS on both sides of PE. The experiments vary the ratio of maneuver capability, the ratio and size of the propellant budgets, the initial separation, the sensor geometry, range, and noise, the game duration, and the defined capture radius. Every configuration is evaluated on the same initial conditions, so that we measure differences between policies on identical encounters. We use this to identify where games are one-sided due to system configuration and where the choice of policy can change the outcome.

The remainder of this paper is organized as follows. \Cref{sec:formulation} presents a general formulation of orbital games and the two games studied. \Cref{sec:orbitalgym} describes the capabilities of OrbitalGym. \Cref{sec:methods} presents the policies studied. \Cref{sec:experiments} describes the experiments and \Cref{sec:results} reports the results. \Cref{sec:conclusions} summarizes the findings and describes future work.

\begin{figure*}[t]
\centering
\includegraphics[width=\textwidth]{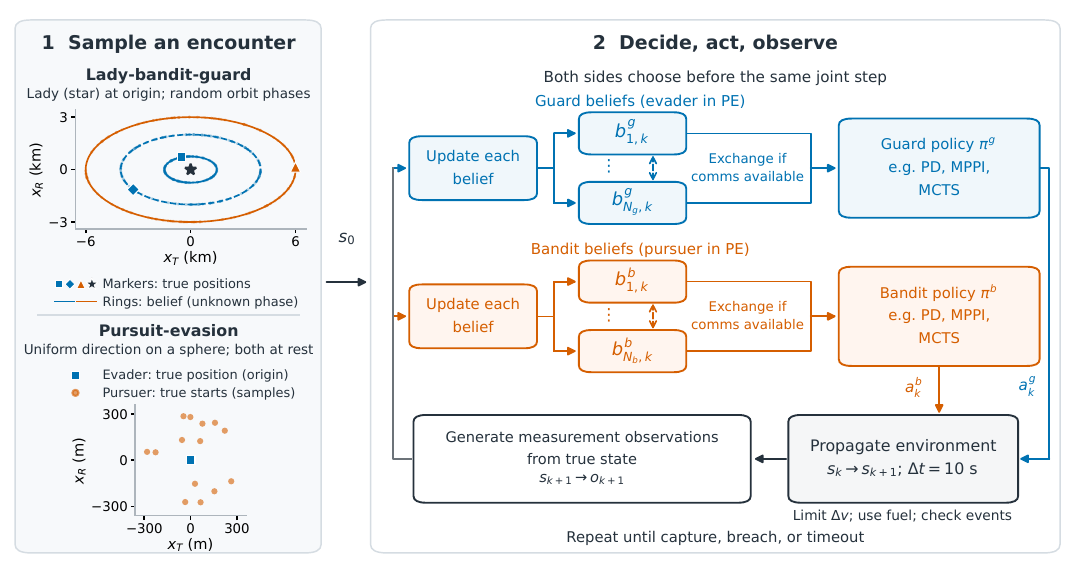}
\caption{Orbital-game setup and decision loop. Left: sampling an encounter. In lady-bandit-guard, one or more guards and bandits start on relative orbits about the lady at randomly drawn phases; in pursuit-evasion, the pursuer starts at the initial separation $d_0$ in a uniformly drawn direction from the evader, both at rest. Right: each vehicle updates its belief from its observations, optionally communicates if possible, the policies of both sides choose commands before the same joint propagation step, and measurements from the new true state feed the next belief update. Dashed arrows indicate optional exchange of beliefs between teammates.}
\label{fig:method-overview}
\end{figure*}

\section{Orbital Games}\label{sec:formulation}
This section defines orbital games generally and the two games that our experiments study in detail: PE and LBG. An orbital game evolves a true state $s_k$ at discrete step $k$ of size $\Delta t$ seconds. Each side controls one or more spacecraft that move relative to a common reference orbit, and each side selects commands based on its own current belief about the game state. The environment applies both sides' commands, propagates the state, generates observations, checks for the events that end the game, and assigns each side a reward $R$. An \emph{episode} is one simulated encounter from its initial state until an ending event or the maximum horizon $T$. \Cref{fig:method-overview} illustrates the loop. Following OrbitalGym's convention, the two sides are the \emph{guard}s, with subscript $G$, and the \emph{bandit}s, with subscript $B$, in every game. In PE the bandit is the pursuer and the guard the evader; in LBG the bandits seek to reach a protected point while the guards seek to prevent them from doing so. We index a side by $i\in\{G,B\}$ and its opponent by $j$.

\subsection{Games}
OrbitalGym provides five games that share the state, dynamics, actions, and observation models below and differ in their objectives and termination rules. \Cref{tab:games} summarizes them. This paper studies pursuit-evasion and lady-bandit-guard~\cite{rusnak2005}. The blocking and minimum separation games assign continuous geometric rewards for remaining near desired locations. \Cref{sec:rewards} gives the rewards of the two games studied.

\begin{table}[t]
\centering\small
\caption{Games provided by OrbitalGym. This paper presents result for pursuit-evasion and lady-bandit-guard.}
\label{tab:games}
\begin{tabular}{@{}p{0.26\columnwidth}p{0.69\columnwidth}@{}}
\toprule
\textbf{Game} & \textbf{Objective} \\
\midrule
Lady-bandit-guard & Guards intercept bandits or keep them from reaching a protected
point, the lady; bandits attempt to reach the lady. \\
\addlinespace
Pursuit-evasion & A pursuer tries to come within the capture radius of an
evader, which tries to avoid it until the horizon. \\
\addlinespace
Sun blocking & Opposing alignment-and-range rewards for positioning between the
opponent and the Sun. \\
\addlinespace
Observation blocking & Alignment-and-range rewards for blocking an Earth target,
gated by target visibility. \\
\addlinespace
Minimum separation & Bandits rewarded for remaining near a protected point while guards
attempts to catch bandits or drive them away. \\
\bottomrule
\end{tabular}
\end{table}

\subsection{Orbit and Attitude Dynamics}
In an Earth-centered inertial frame, the position $\mathbf p$ of a spacecraft obeys the nonlinear two-body equation with perturbing accelerations $\mathbf a_{\mathrm{p}}$,
\begin{equation}
\ddot{\mathbf p}=-\frac{\mu}{\|\mathbf p\|^3}\,\mathbf p+\mathbf a_{\mathrm{p}}
\label{eq:two-body}
\end{equation}
where $\mu$ is the gravitational parameter of Earth. OrbitalGym propagates \Cref{eq:two-body} by numerical integration for unperturbed motion and configurable force models that optionally add spherical-harmonic gravity, atmospheric drag, solar radiation pressure, and third-body attraction through the Astrojax library~\cite{astrojax}.

An orbital game concerns the vehicles' positions relative to one another and to a protected point, not their inertial positions, so we express the game relative to a reference point on a circular orbit of radius $a_{\mathrm{ref}}$ with mean motion $n=\sqrt{\mu/a_{\mathrm{ref}}^3}$. In LBG the lady is this reference point. Positions are given in the rotating radial, along-track, and cross-track (RTN) frame centered at the reference point. For separations that are small compared with $a_{\mathrm{ref}}$, the motion of a vehicle with RTN position $(x_R,x_T,x_N)$ follows the HCW equations
\begin{equation}
\begin{aligned}
\ddot x_R&=3n^2x_R+2n\dot x_T\\
\ddot x_T&=-2n\dot x_R\\
\ddot x_N&=-n^2x_N
\end{aligned}
\label{eq:hcw}
\end{equation}
between maneuvers, which linearize the unperturbed relative two-body dynamics about the circular reference orbit~\cite{hill1878,clohessy1960}. Let $s=[x_R,x_T,x_N,\dot x_R,\dot x_T,\dot x_N]^\top$ be the relative orbital state of a vehicle, so that \Cref{eq:hcw} reads $\dot s=Fs$ with a constant matrix $F$. Its solution over an interval $\tau$ is $s(t+\tau)=A(\tau)s(t)$, where the state-transition matrix $A(\tau)=e^{F\tau}$ has a closed form. If the vehicle applies an impulsive velocity change $u_k$ at the start of step $k$, the discrete-time state update is
\begin{equation}
s_{k+1}=A(\Delta t)\left(s_k+Bu_k\right),\qquad
B=\begin{bmatrix}0_{3\times3}\\I_3\end{bmatrix}
\label{eq:transition}
\end{equation}
where the input matrix $B$ adds the impulse to the velocity components. \Cref{eq:transition} is the exact discrete-time solution of \Cref{eq:hcw} under the impulsive maneuver model. It lets a planner predict the consequences of a sequence of maneuvers far into the future with one matrix product per step, which is what makes sampling-based planning over hundreds of candidate sequences computationally feasible. Because \Cref{eq:hcw} is linear, the state of the bandit relative to the guard, $x_k=s_{B,k}-s_{G,k}$, obeys the same transition with the difference of the two impulses
\begin{equation}
x_{k+1}=A(\Delta t)\left(x_k+B\left(u_{B,k}-u_{G,k}\right)\right)
\label{eq:relative}
\end{equation}
We write $r_k$ for the first three components of $x_k$, the relative position, and $v_k$ for the last three, the relative velocity.

For some games it is also important to consider the vehicle attitude because the sensors and radios that provide a vehicle's information are fixed to its body: a camera measures relative position only within a cone about its boresight, and a directional antenna communicates only toward where it points. When included in a game formulation, OrbitalGym represents attitude by a unit quaternion $q$ and body rate $\omega$ and provides two attitude dynamics models. The rigid-body model integrates Euler's equations with an inertia tensor under a commanded torque. A simplified kinematic model points the body towards a commanded direction, rotating by at most the slew rate times $\Delta t$ in one step.

\subsection{Thruster Modeling}
Although occasionally omitted from control-oriented studies of spacecraft guidance, propellant is a central consideration in an orbital game, because a vehicle that has spent its propellant can no longer maneuver, and a planner that does not account for propellant use cannot reason about how its expenditure affects the outcome. OrbitalGym can extend the vehicle state with its mass. A vehicle has dry mass $m_d$, propellant mass $m_{p,k}$ at step $k$, wet mass $m_k=m_d+m_{p,k}$, maximum thrust $F$, and specific impulse $I_{sp}$. Two limits bound the maximum velocity change possible in one step: the thrust acting over the fixed step $\Delta t$ and the propellant that remains
\begin{equation}
\Delta v_k^{\max}=\min\left\{\frac{F\Delta t}{m_k},\;I_{sp}g_0\ln\frac{m_k}{m_d}\right\}
\label{eq:dvmax}
\end{equation}
where $g_0$ is standard gravity. A commanded impulse $u_k^{\mathrm{cmd}}$ is scaled down along its direction to this limit, $u_k=u_k^{\mathrm{cmd}}\min\{1,\Delta v_k^{\max}/\|u_k^{\mathrm{cmd}}\|\}$. The rocket equation then gives the propellant consumed
\begin{equation}
m_{p,k+1}=m_{p,k}-m_k\left[1-\exp\left(-\frac{\|u_k\|}{I_{sp}g_0}\right)\right]
\label{eq:rocket}
\end{equation}
Under this model the velocity change available per step grows as the vehicle lightens, and the total velocity change available over a game is fixed by the wet-to-dry mass ratio, $I_{sp}g_0\ln(m_0/m_d)$.

A simpler actuation model holds the mass constant, which separates the limit on single-step maneuverability from the limit on total maneuverability so that each can be varied on its own. Each side $i$ has a \emph{thrust limit} $\bar u_i$, the maximum velocity change in one step, and a \emph{delta-v budget} $\Delta v_i$, the maximum total velocity change over the game,
\begin{equation}
\|u_{i,k}\|_2\le\bar u_i,\qquad
\sum_{k}\|u_{i,k}\|_2\le\Delta v_i
\label{eq:limits}
\end{equation}
which is \Cref{eq:dvmax,eq:rocket} with the mass held constant. Specifying $\Delta v_i$ directly is equivalent to specifying the propellant mass, as the budget is the usual form of the constraint, and $\bar u_i$ corresponds to $F\Delta t/m$. We describe the relative capability of the two sides with the \emph{thrust ratio} $\rho_u$, the ratio of the maximum single-step velocity changes, and the \emph{budget ratio} $\rho_{\Delta v}$, the ratio of the total velocity changes allowed over the game. In both games the ratio divides the guard's capability by the bandit's,
\begin{equation}
\rho_u=\frac{\bar u_G}{\bar u_B},\qquad
\rho_{\Delta v}=\frac{\Delta v_G}{\Delta v_B}
\label{eq:ratios}
\end{equation}
so a ratio above one favors the guard and a ratio below one favors the bandit, whichever side pursues.

\subsection{Observations and Beliefs}
A side acts on its \emph{belief}, an estimate of the state of the other vehicles along with its uncertainty about that state. Under \emph{full information} the belief is the true relative state. Otherwise, each side measures the relative position each step and uses those measurements to update its belief. We write the measurement of either side in the convention of \Cref{eq:relative}, the bandit's position relative to the guard,
\begin{equation}
\begin{aligned}
z_{i,k}&=r_k+\epsilon_{i,k},\qquad
\epsilon_{i,k}\sim\mathcal N\left(0,\sigma_k^2I_3\right)\\
\sigma_k&=\max\left(\eta\|r_k\|_2,\,\qty{0.1}{\meter}\right)
\end{aligned}
\label{eq:measurement}
\end{equation}
where the noise fraction $\eta$ makes the noise proportional to range. A bandit that senses the guard relative to itself measures $-r_k$; negating it gives \Cref{eq:measurement}, and the isotropic noise is unchanged. We consider two types of sensors. A \emph{radial sensor}, used in some game formulations, detects the other vehicle in every direction within the sensor range $d_s$. A \emph{cone sensor} mimics a body-fixed camera, which sees only within a cone about its boresight. Let $r_i$ and $r_j$ be the RTN positions of the observer and the target, and $\hat b_i$ the observer's unit boresight, which is fixed in its body frame and rotated into the RTN frame by its attitude $q_i$. With cone half-angle $\theta_s$, the target is observed when it lies within $d_s$ and
\begin{equation}
\hat b_i^\top\frac{r_j-r_i}{\|r_j-r_i\|}\ge\cos\theta_s
\label{eq:cone}
\end{equation}

Each side estimates the relative state with an Extended Kalman filter~\cite{simon2006optimal} or particle filter~\cite{thrun2002probabilistic}. A side knows its own impulse but cannot observe the opponent's directly. The filter predicts the next state with \Cref{eq:relative} using its own impulse and treats the opponent's impulse $u_j$ as unknown, with zero mean and covariance $\bar u_j^2I_3$, where $\bar u_j$ is the opponent's thrust limit. Because the impulse enters through $B$ before the step is propagated, the estimate covariance $P$ is predicted as $A(P+\bar u_j^2BB^\top)A^\top$ with $A=A(\Delta t)$, so the process-noise covariance in state space is $\bar u_j^2AB B^\top A^\top$. The initial belief, or \emph{prior}, is the true state perturbed by Gaussian errors with standard deviations $\sigma_r$ in position and $\sigma_v$ in velocity.

\subsection{Communication}
When a side has more than one vehicle, its teammates may communicate  to exchange beliefs to jointly plan actions. OrbitalGym evaluates a \emph{link} model every step, which decides which teammates are in contact. The available link models are: always linked, linked within a maximum distance, linked while a teammate lies inside a body-fixed communication cone, either one-sided or mutual as for a laser crosslink, or linked while the reference orbit is within view of a ground-station network. Teammates in contact fuse their beliefs about each opposing vehicle, by information-form pooling for Kalman filters and by joint resampling for particle filters, while teammates out of contact keep their own estimates. Communicating can also be an explicit action with a potential cost: in LBG, a transmission can reveal the communicator's position. While communications modeling is not exercised in the experiments of this paper due to only considering single-agent scenarios, the capability is included in OrbitalGym to support the study of multi-agent scenarios.

\subsection{Captures, Outcomes, and Rewards}\label{sec:rewards}
Two events end a game. A \emph{capture} occurs when the pursuer in PE, or the guard in LBG, comes within the capture radius $r_c$ of its opponent. A capture may also carry a \emph{velocity-matching condition}: the relative speed must satisfy $\|v_k\|\le v_e$ at the moment of capture, where $v_e$ is the largest relative speed that counts as matched. When $v_e=\infty$ position alone defines the capture. A \emph{breach} occurs in LBG when the bandit comes within the breach radius $r_b$ of the lady, which sits at the origin of the frame, subject to the same velocity-matching condition if applicable. In PE the pursuer wins by capture and the evader by avoiding it until the horizon $T=K\Delta t$; survival to $T$ does not mean that the evader escaped permanently. In LBG the guard wins by capture or by keeping the bandit out until $T$, and the bandit wins by breach. When both events occur in the same step, the event that occurs earlier within the step decides, and a tie counts as a capture. We report the \emph{capture fraction} in PE and the \emph{defense fraction} in LBG, the fractions of encounters that the pursuer or the guard wins.

The outcome of an episode is a binary win or loss for each side. The rewards that the planners optimize are denser than the outcome. A planner at step $k$ evaluates candidate maneuver sequences by predicting the encounter $h=1,2,\dots$ steps ahead and summing a stage reward over those lookahead steps. Let $\tilde r_{k+h}=\|r_{k+h}\|/\|r_k\|$ be the predicted separation of the two vehicles $h$ steps ahead, normalized by the separation at the current step $k$, $\tilde r^{L}_{k+h}$ the bandit's predicted distance to the lady normalized in the same way, and $b$ an event bonus. The indicators $\delta^{c}_{k+h}$ and $\delta^{b}_{k+h}$ equal one if a capture or a breach is predicted in the step that ends at $k+h$ and zero otherwise. The stage reward of side $i$ for lookahead step $h$ is
\begin{equation}
\begin{aligned}
\text{PE:}\quad R_{B,k+h}&=-\tilde r_{k+h}+b\,\delta^{c}_{k+h}\\
R_{G,k+h}&=-R_{B,k+h}\\
\text{LBG:}\quad R_{G,k+h}&=-\tilde r_{k+h}+b\left(\delta^{c}_{k+h}-\delta^{b}_{k+h}\right)\\
R_{B,k+h}&=-\tilde r^{L}_{k+h}+b\left(\delta^{b}_{k+h}-\delta^{c}_{k+h}\right)
\end{aligned}
\label{eq:rewards}
\end{equation}
The denominator is the separation at the step the plan starts from, floored at the capture radius (the breach radius for $\tilde r^{L}$), neither the separation at the end of the lookahead nor the initial separation $d_0$ of the episode. In the step of an event the side receives the bonus alone, and nothing after it. Because a planner replans at every step, it is renewed at every step from the current belief, which keeps the distance term near one, and so comparable with $b$, at any range. In PE the bandit is rewarded for closing on the guard and the guard receives the negative; in LBG the guard is rewarded for closing on the bandit and the bandit for closing on the lady. When an event has a velocity-matching condition, the side that must satisfy it receives an additional penalty equal to the relative speed in excess of the glideslope $v_e+\|r\|/\tau_a$, divided by $v_e$, with $\tau_a=\qty{100}{\second}$, which rewards slowing down as it approaches. %The planners minimize the per-side stage cost $c_i=-R_i$.

The distance terms of \Cref{eq:rewards} are potentials, functions of the state alone that rise as a side nears its objective: $\phi_G(s)=-\tilde r$ for a guard closing on a bandit and $\phi_B(s)=-\tilde r^{L}$ for a bandit closing on the lady. \Cref{eq:rewards} pays the potential itself at every step, which suits a planner that looks a fixed number of steps ahead and may not see the event that ends the game. OrbitalGym's native LBG reward, intended for learning, instead pays the change in a potential on top of the event bonuses, with an optional term in the guard's potential for its distance to the lady. For weight $\alpha$ and discount $\gamma$, the shaping reward of a transition from $s$ to $s'$ is
\begin{equation}
F_i(s,s')=\alpha\left[\gamma\phi_i(s')-\phi_i(s)\right]
\label{eq:shaping}
\end{equation}
Summed over an episode, these differences add up to a quantity that depends only on the initial state, so the shaping guides a learner toward its objective without changing which policy is optimal against a fixed opponent~\cite{ng1999shaping}.

\begin{figure*}[t]
\centering
\includegraphics[width=\textwidth]{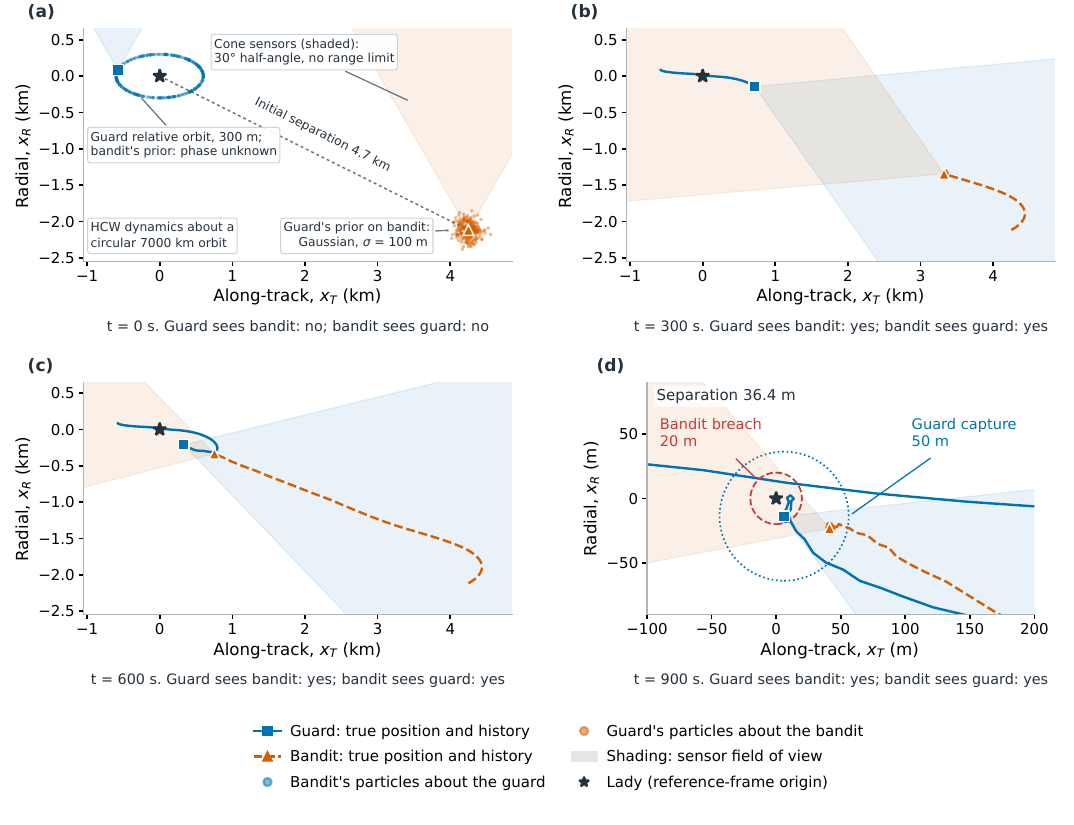}
\caption{An illustrative LBG encounter under OrbitalGym's reference configuration, in which the guard captures the bandit at \qty{900}{\second}. Panel (a) labels the modeling choices that define the encounter: the initial separation, the guard's relative orbit, each side's prior, the conical sensors, and the dynamics. Blue denotes the guard and orange the bandit; particles show each side's belief about the other and curves show true trajectories. Shading shows the sensor cones intersected with the plotted plane. This reference configuration uses particle filters, \ang{30} cones, a \qty{50}{\meter} capture radius, and a \qty{20}{\meter} breach radius with a repeated-proximity requirement; panel (d) shows a close-up of the game-ending event conditions.}
\label{fig:game-progression}
\end{figure*}

\Cref{fig:game-progression} illustrates sensing, belief, and control in one LBG encounter under OrbitalGym's reference LBG configuration, and labels the modeling choices that define an encounter.

\begin{lstlisting}[ float=tbp, caption={Pursuit-evasion in OrbitalGym. The guard takes random impulses, the bandit plans with the library's MPPI policy against a forecast of a coasting guard, and one compiled call rolls out 1,024 encounters.}, label={lst:example}, ]
from types import SimpleNamespace as View

import jax
import jax.numpy as jnp
import orbitalgym as og
from orbitalgym import policies as pol
from orbitalgym.rollout import rollout

keys = jax.random.split(jax.random.key(0), 1024)
cfg = og.make_pursuit_evasion(
    capture_distance_m=10.0)
env = og.OrbitalGymEnv(cfg)

# Guard: random 0.25 m/s impulses along RTN axes.
axes = jnp.concatenate([jnp.eye(3), -jnp.eye(3)])
guard = pol.UniformRandomDiscretePolicy(
    0.25 * axes, 1, env.guard_command_cls)

# Bandit: MPPI planning against a coasting guard.
coast = pol.ZeroControl(env.guard_command_cls, 1)
mppi = pol.MPPIPolicy(
    og.POMDPAdapter(env), og.Side.BANDIT, coast,
    n_samples=64, horizon=10, dv_max=0.5,
    n_vehicles=1,
    command_cls=env.bandit_command_cls,
    template_env_state=env.reset(keys[0])[0])

# rollout gives each policy its flat observation;
# MPPI reads it as an estimate of both vehicles.
def bandit(plan, obs, key, t):
    view = View(mean=obs.reshape(1, 2, 6))
    return mppi(plan, view, key, t)

policies = og.BySide(guard=guard, bandit=bandit)
# Each side's initial policy state: MPPI's nominal
# impulse sequence; None for the stateless guard.
init_state = og.BySide(
    guard=lambda *_: None,
    bandit=lambda *_: mppi.init_state())

@jax.jit
@jax.vmap
def episode(key):
    s = rollout(env, policies, init_state, key,
                cfg.max_steps).final_state
    d = s.guards.rtn[0, :3] - s.bandits.rtn[0, :3]
    return jnp.linalg.norm(d), s.step * cfg.dt

dist, t_end = episode(keys)
caught = dist < cfg.game.capture_distance_m
t_mean = t_end[caught].mean()
print(f"capture rate {caught.mean():.2f}, "
      f"mean capture time {t_mean:.0f} s")
\end{lstlisting}

\section{OrbitalGym}\label{sec:orbitalgym}
OrbitalGym is an environment for defining, simulating, and solving orbital games. It is written in JAX~\cite{jax2018}, so that a game transition is a stateless-function acting on input arrays that can be compiled, vectorized over many independent episodes, and executed on a graphics processing unit (GPU) in parallel. This section describes its components and how an experiment uses them.

OrbitalGym separates an orbital game into components that can be composed and varied independently. \Cref{tab:components} lists them. The \emph{truth dynamics} propagate the state of every vehicle. The orbit state can be extended with the attitude dynamics and fuel use. Vehicle state can include position and velocity,  attitude, body rates, and propellant mass. A side's planning model, observation model and belief model are separate components, so a planner can use a dynamics model that differs from the truth, such as using HCW dynamics for planning and numerical integration for belief updates. Teams of more than one vehicle exchange beliefs over communication links, as described in \Cref{sec:formulation}. The game variant (e.g. PE, LBG) defines the reward and termination rules. 

\begin{table}[t]
\centering\small
\caption{OrbitalGym components.}
\label{tab:components}
\begin{tabular}{@{}p{0.28\columnwidth}p{0.66\columnwidth}@{}}
\toprule
\textbf{Component} & \textbf{Provided implementations} \\
\midrule
Orbit dynamics & HCW (closed form); two-body, $J_2$, and configurable force models (numerically integrated) \\
Attitude dynamics & Rigid body (numerically integrated), kinematic pointing \\
Vehicle state & Position, velocity, propellant mass, attitude, body rate \\
Actuation & Impulse cap, rocket equation, torque and slew limits \\
Observations & Full, range-limited, conical, onboard navigation (GPS) \\
Beliefs & Kalman, extended Kalman, and particle filters \\
Communication & Links by distance, pointing cone, or ground-station network; belief fusion between linked teammates; communication as an observable action \\
Policies & Non-reactive (coast, random); classical feedback control (PD and LQR); MPPI; Monte Carlo tree search.  \\
Interfaces & Native JAX, Gymnasium, PettingZoo \\
\bottomrule
\end{tabular}
\end{table}

A game advances by one step through a single function that takes the state and both sides' commanded actions and returns the next state, the observations, the rewards, and a termination flag. Because the function is stateless, \texttt{jax.vmap} evaluates it on a array of input states and \texttt{jax.lax.scan} rolls it out over a game. \Cref{lst:example} builds a PE game, gives each side a policy that ships with OrbitalGym, and rolls out 1,024 encounters in one compiled call. The Gymnasium and PettingZoo adapters wrap the same step function for single-agent and multi-agent learning libraries.

The experiments in this paper run every episode through the same step function. Each experiment compiles one program per combination of policy types and evaluates many configurations, each on the same initial conditions, in one call. On one NVIDIA H100 GPU, a batch of 11,520 encounters of 180 steps with a \qty{10}{\second} step takes about \qty{1}{\second} with feedback control policies on both sides, \qty{28}{\second} with an MPPI planner on one side, and \qty{55}{\second} with MPPI planners on both sides. Each run records the configuration, the seeds that determined the initial conditions, the source revision, and one row per episode, and scripts regenerate the result figures and the quoted statistics from these records.

\section{Policies}\label{sec:methods}
This section introduces the policies that the experiments compare. We choose policies that span a range of decision-making effort: policies that do not react to the opponent, feedback control laws that react to the current state, and planners that optimize over forecasts of the future. \Cref{tab:policies} summarizes them, and \Cref{app:controllers} gives the full formulation of each, including the gains, weights, and planner settings. Every policy acts on the side's belief, so a policy that is written in terms of the relative state $x$ uses the estimate $\hat x$ for evaluation, and every commanded action is clipped by \Cref{eq:limits} before it is applied.

\begin{table*}[t]
\centering\small
\caption{Policies. The PE pursuer and the LBG guard use the same feedback laws on their relative state to the opponent; the LBG bandit steers toward the lady.}
\label{tab:policies}
\begin{tabular}{@{}lllp{0.5\textwidth}@{}}
\toprule
\textbf{Game} & \textbf{Side} & \textbf{Policy} & \textbf{Decision rule} \\
\midrule
PE and LBG & Pursuer, guard & Coast & No impulse \\
 & & Direct & Full impulse along the line of sight to the opponent \\
 & & PD & Proportional-derivative feedback on the relative position and velocity \\
 & & LQR & Linear-quadratic regulator on the relative state \\
 & & HCW intercept & Impulse that reaches the opponent after a lookahead $T_L$ if it coasts; $T_L=100$, 300, or \qty{600}{\second} \\
 & & MPPI & MPPI forecasting a coasting opponent (MPPI-Coast) or one that flees or approaches the lady (MPPI-Flee, MPPI-Approach) \\
 & & MCTS & PE only. Tree search over nine impulses held for \qty{30}{\second} (coast; full thrust along $\pm$R, $\pm$T, $\pm$N; full thrust toward the evader; clipped HCW intercept). Evader policy drawn at each edge from \{coast, flee\}. Leaves valued by rollouts of HCW intercept \\
\midrule
PE & Evader & Coast, random & No impulse; uniform random impulse \\
 & & Flee, transverse & Full impulse away from, or normal to, the line of sight \\
 & & MPPI-Direct, MPPI-LQR & MPPI forecasting direct or LQR pursuit \\
 & & MCTS & Tree search over nine impulses held for \qty{30}{\second} (coast; full thrust along $\pm$R, $\pm$T, $\pm$N; full thrust away from the pursuer; transverse burn). Pursuer policy drawn at each edge from \{coast, direct, HCW intercept\}. Leaves valued by rollouts of the fleeing policy \\
\midrule
LBG & Bandit & Coast, direct & No impulse; full impulse toward the lady \\
 & & LQR, HCW intercept & Regulator or intercept to the lady \\
 & & Evade & Toward the lady, turning away from a guard within \qty{300}{\meter} \\
 & & MPPI-Direct, MPPI-LQR & MPPI forecasting direct or LQR guard pursuit \\
\bottomrule
\end{tabular}
\end{table*}

\subsection{Non-Reactive and Feedback Control Policies}
The \emph{coast} policy applies no impulse and the \emph{random} policy applies an impulse drawn uniformly from the ball of admissible impulses at every step; they serve as reference baselines. The \emph{direct} policy applies full thrust along the line of sight to, or away from, its target with no velocity term, so it can overshoot a stationary target. The \emph{PD} policy applies proportional and derivative feedback on the relative position and velocity. The \emph{LQR} policy applies $u=-Kx$, where $K$ solves the discrete-time infinite-horizon regulator problem for \Cref{eq:relative} with the opponent's impulse set to zero. The \emph{HCW intercept} policy computes the single impulse that brings the relative position to zero after the lookahead $T_L$ if the opponent coasts, and recomputes it at every step. The pursuer in PE and the guard in LBG apply these laws to their opponent-relative state, and the LBG bandit applies direct, LQR, and HCW intercept to its position relative to the lady.

The PE evader has two reactive policies. The \emph{flee} policy applies full thrust directly away from the pursuer; in Wong's constant-gravity, constant-thrust, propellant-limited games, thrusting in a fixed direction is the optimal evasion strategy~\cite{wong1967}, which motivates it. The \emph{transverse} policy applies full thrust normal to the line of sight, which forces the pursuer to turn. The LBG bandit's \emph{evade} policy steers toward the lady and turns away from a guard that comes within \qty{300}{\meter}.

\subsection{Model Predictive Path Integral Planning}
MPPI~\cite{williams2017mppi} is a sampling-based model predictive controller. At every step, it samples $N$ impulse sequences around a nominal sequence, simulates each over a horizon of $H$ steps with a model of the game and a forecast $\pi_j$ of the opponent's policy, weights each sequence by the exponential of its negative cost, and moves the nominal sequence to the weighted average. The side executes the first impulse and shifts the nominal sequence by one step for the next decision. The nominal sequence that starts the game is zero-thrust control, that is, coasting, and we include the zero sequence and the current nominal among the samples, so that the planner can always choose to coast or to continue its plan. The costs are the stage costs of \Cref{sec:rewards}.

The planner needs a forecast of the opponent actions and we name each MPPI variant by the policy it assumes its opponent follows. In PE, the pursuer's \emph{MPPI-Coast} forecasts a coasting evader and \emph{MPPI-Flee} one that flees at full thrust; the evader's \emph{MPPI-Direct} forecasts a pursuer that thrusts directly at it and \emph{MPPI-LQR} the LQR pursuer. In LBG, the guard's \emph{MPPI-Coast} forecasts a coasting bandit and \emph{MPPI-Approach} one that thrusts directly at the lady; the bandit's \emph{MPPI-Direct} and \emph{MPPI-LQR} forecast direct and LQR guards. All MPPI policies use the same default settings, listed in \Cref{app:mppi} with the full procedure, which we did not tune per configuration; the lookahead and sample-count experiments vary them deliberately.

\subsection{Monte Carlo Tree Search}
MCTS~\cite{kocsis2006uct,browne2012mcts} builds a search tree over discrete actions by repeated simulation. Our search, implemented for PE on the mctx library~\cite{mctx}, combines three components: the set of actions that the tree searches, the set of opponent policies from which the opponent's behavior is drawn, and a rollout policy that values each new leaf. The tree chooses among nine impulses, each held for three steps: coasting, full thrust along each of the six RTN axes, full thrust along the line of sight, and the clipped HCW intercept for the pursuer or, for the evader, the transverse policy's full-thrust burn perpendicular to the line of sight. The side does not know the opponent's policy, so, following Stochastic MuZero~\cite{antonoglou2022stochastic}, every edge passes through a chance node that draws the opponent's policy uniformly from a set: coasting or fleeing for the pursuer's search, and coasting, direct pursuit, or HCW intercept for the evader's. The value of a simulated path is the discounted sum of the stage rewards of \Cref{sec:rewards} collected along its tree edges, plus the discounted value of its leaf; a new leaf is valued by a Monte Carlo rollout to the end of the lookahead in which the searching side follows the rollout policy, HCW intercept for the pursuer and fleeing for the evader, against an opponent policy drawn from the same set. The rollout policy only values leaves; the side's actions always come from the nine impulses of the tree, and the side executes the most visited one. \Cref{app:mcts} gives the settings.

\section{Experiments}\label{sec:experiments}
Now that we have a flexible simulation environment able to simulate a wide range of scenarios with different satellite system configurations, we use it to run a range of experiments to help characterize where the choice of policy changes the outcome and where the outcome is set by capability regardless of policy. This section describes the shared setup and the experiments.

We use four terms throughout. A \emph{configuration} fixes everything about an encounter except where it starts: the game, the policy of each side, and the scenario parameters, that is, the physical parameters, constraints, the budgets, the distribution of initial states, the horizon, the sensor, and the event definition. An \emph{initial condition} is one random draw of what the configuration leaves open: the state in which each vehicle starts and the error of its initial belief. An \emph{encounter} is one simulation of a configuration from one initial condition, with its own random seed for measurement noise, random policies, and the planners' sampling. An \emph{experiment} is a set of configurations that varies some parameters and holds the others fixed. For every configuration we simulate 128 encounters and report the fraction that the pursuer wins in PE (the capture fraction) or the guard wins in LBG (the defense fraction). Every configuration of an experiment uses the same 128 initial conditions and seeds, so two configurations are compared on the same encounters.

Both games have one vehicle on each side, so no communication occurs and each side's belief is its own. The truth dynamics are the HCW model about a circular reference orbit of radius \qty{7000}{\kilo\meter} with a step of $\Delta t=\qty{10}{\second}$, actuation follows the constant-mass limits of \Cref{eq:limits}, and every event is tested at $S=10$ points within each step. The pursuing side, the bandit in PE ($\bar u_B$, $\Delta v_B$) and the guard in LBG ($\bar u_G$, $\Delta v_G$), has a thrust limit of $\bar u_B=\bar u_G=\qty{0.1}{\meter\per\second}$ per step, equivalent to an acceleration of \qty{0.01}{\meter\per\second\squared}, and unless stated a delta-v budget of $\Delta v_B=\Delta v_G=\qty{5}{\meter\per\second}$; the ratios of \Cref{eq:ratios} then set the other side's thrust limit and delta-v budget. In PE the evader starts at the reference point and the pursuer at the initial separation $d_0$, by default \qty{300}{\meter}, in a direction drawn uniformly on the sphere, both at rest in the rotating frame; the capture radius is $r_c=\qty{10}{\meter}$. In LBG the bandit starts \qty{1000}{\meter} from the lady and the guard \qty{200}{\meter} from the lady, each in an independent uniform direction and at rest; the capture radius is $r_c=\qty{10}{\meter}$ and the breach radius $r_b=\qty{50}{\meter}$. We choose $r_c<r_b$ so that a guard that waits at the lady cannot win by default. Unless stated the horizon is $T=\qty{1800}{\second}$ and events have no velocity-matching condition.

The baseline gives both sides full information. We choose this baseline deliberately: it removes estimator error, so that a difference between two configurations that share geometry, horizon, and event definition can be attributed to capability or to the policy. The information experiments then remove it, including a repetition of the PE capability experiment under a cone sensor. There, each side estimates the relative state with a Kalman filter from the measurements of \Cref{eq:measurement} taken by a radial or a cone sensor, and a cone sensor's boresight points along the estimated direction to the opponent at every step, the kinematic attitude model with an unlimited slew rate. A cone sensor can still lose its target when the target is beyond the sensor range or when the direction of the estimate is wrong by more than the cone half-angle.

\Cref{tab:experiments,tab:experiments-lbg} list the PE and LBG experiments, the parameters each varies, and the parameters each holds fixed. The policy gain compares two policies on the same 128 encounters. We report win fractions and policy gains as percentages, so a policy gain of 10\% is a difference of 10 percentage points between two win fractions. We count a gain as meaningful when its magnitude is at least 10\%, and every gain counted in \Cref{sec:results} differs from zero at the 95\% confidence level based on the observed standard error. An independent repetition of both capability experiments with 512 new encounters per configuration confirmed the policy gain findings: the win fractions of the two experiments differ by 0.3\% on average in PE and 0.9\% in LBG, and the number of configurations with a meaningful policy gain changes from 15 to 17 of 90 against the MPPI-LQR evader and from 74 to 75 against the MPPI-LQR bandit.

\begin{table*}[!t]
\centering\small
\sisetup{list-final-separator={, }, list-pair-separator={, }}
\caption{Pursuit-evasion experiments, with the identifiers that the figure captions cite. Each experiment evaluates the listed values for the listed pursuers and evaders, or guards and bandits, with the stated parameters held constant. Experiments are fully crossed except where a row names specific combinations. Policy names follow \Cref{tab:policies}. Operating points are pairs $(\rho_u,\rho_{\Delta v})$; unless stated, $\bar u_B=\qty{0.1}{\meter\per\second}$, $\Delta t=\qty{10}{\second}$, $S=10$, and both sides have full information.}
\label{tab:experiments}
\begin{tabular}{@{}>{\raggedright\arraybackslash}p{0.15\textwidth}>{\raggedright\arraybackslash}p{0.16\textwidth}>{\raggedright\arraybackslash}p{0.39\textwidth}>{\raggedright\arraybackslash}p{0.228\textwidth}@{}}
\toprule
\textbf{Experiment} & \textbf{Varied parameter} & \textbf{Values} & \textbf{Held constant} \\
\midrule
PE-1: capability & Thrust ratio $\rho_u$ & 0.25, 0.5, 0.75, 0.9, 1, 1.1, 1.25, 1.5, 2 & \multirow[t]{4}{=}{$d_0=\qty{300}{\meter}$, $T=\qty{1800}{\second}$, $\Delta v_B=\qty{5}{\meter\per\second}$, $r_c=\qty{10}{\meter}$, no velocity-matching condition} \\
 & Budget ratio $\rho_{\Delta v}$ & 0.25, 0.5, 0.75, 0.9, 1, 1.1, 1.25, 1.5, 2, $\infty$ &  \\
 & Pursuer & Coast, direct, PD, LQR, HCW intercept ($T_L=\qty{100}{\second}$), MPPI-Coast, MPPI-Flee &  \\
 & Evader & Coast, random, flee, transverse, MPPI-Direct, MPPI-LQR &  \\
\addlinespace
PE-2: budget size & Pursuer budget $\Delta v_B$ & \qty{1}{\meter\per\second}, \qty{20}{\meter\per\second}, each over the capability sweep; unlimited for both sides over $\rho_u$ & \multirow[t]{3}{=}{$d_0=\qty{300}{\meter}$, $T=\qty{1800}{\second}$, $r_c=\qty{10}{\meter}$, no velocity-matching condition} \\
 & Pursuer & Coast, direct, PD, LQR, HCW intercept ($T_L=\qty{100}{\second}$), MPPI-Coast, MPPI-Flee &  \\
 & Evader & Coast, random, flee, transverse, MPPI-Direct, MPPI-LQR &  \\
\addlinespace
PE-3: geometry & Initial separation $d_0$ & \qtylist{100;200;400;800;1600;3200}{\meter} & \multirow[t]{5}{=}{$\Delta v_B=\qty{5}{\meter\per\second}$, $r_c=\qty{10}{\meter}$, no velocity-matching condition} \\
 & Horizon $T$ & \qtylist{300;600;900;1800;3600;5400}{\second} &  \\
 & Operating point & (1, 0.75), (0.5, 1), (1, 0.5) &  \\
 & Pursuer & Coast, direct, PD, LQR, HCW intercept ($T_L=\qty{100}{\second}$), MPPI-Coast, MPPI-Flee &  \\
 & Evader & Coast, random, flee, MPPI-LQR &  \\
\addlinespace
PE-4: information & Sensor & Full; radial; conical with $\theta_s=\ang{10}$ & \multirow[t]{6}{=}{$T=\qty{1800}{\second}$, $\Delta v_B=\qty{5}{\meter\per\second}$; with a sensor, both sides sense and start from a prior with $\sigma_r=0.1d_0$, $\sigma_v=\qty{0.01}{\meter\per\second}$} \\
 & Sensor range $d_s$ & \qty{0.5}{\kilo\meter}, \qty{5}{\kilo\meter} &  \\
 & Noise fraction $\eta$ & 1\% (both ranges); 10\% for the radial sensor at \qty{5}{\kilo\meter} &  \\
 & Initial separation $d_0$ & \qty{300}{\meter}, \qty{1000}{\meter} &  \\
 & Operating point & (1, 0.75), (0.5, 1), (1, 0.5) &  \\
 & Pursuer & Coast, direct, PD, LQR, HCW intercept ($T_L=\qty{100}{\second}$), MPPI-Coast, MPPI-Flee &  \\
 & Evader & Coast, random, flee, MPPI-LQR &  \\
 & Capability sweep & Every ratio, pursuer, and evader of PE capability under the \qty{5}{\kilo\meter} cone sensor with 1\% noise, $d_0=\qty{300}{\meter}$ &  \\
\addlinespace
PE-5: lookahead & MPPI-Flee lookahead & \qtylist{90;210;360;600;900}{\second} & \multirow[t]{3}{=}{Operating point (1, 0.75), $d_0=\qty{300}{\meter}$, $T=\qty{1800}{\second}$} \\
 & Pursuer information & Full; radial \qty{5}{\kilo\meter} with $\eta=0.1\%$ or 1\% (pursuer only) &  \\
 & Evader & Flee, random, MPPI-LQR (lookahead \qty{600}{\second}) &  \\
\addlinespace
PE-6: evader forecast & Evader & MPPI-Direct, MPPI-LQR & \multirow[t]{4}{=}{$\rho_{\Delta v}=1$, $d_0=\qty{300}{\meter}$, $T=\qty{1800}{\second}$, $\Delta v_B=\qty{5}{\meter\per\second}$} \\
 & Evader samples & 64, 256, 1024 &  \\
 & Thrust ratio $\rho_u$ & 0.5, 0.75, 1 &  \\
 & Pursuer & LQR, HCW intercept, MPPI-Coast, MPPI-Flee &  \\
\addlinespace
PE-7: capture definition & Capture radius $r_c$ & \qtylist{5;10;20;50}{\meter} & \multirow[t]{5}{=}{$d_0=\qty{300}{\meter}$, $T=\qty{1800}{\second}$, $\Delta v_B=\qty{5}{\meter\per\second}$} \\
 & Velocity matching $v_e$ & None, \qty{2}{\meter\per\second}, \qty{0.5}{\meter\per\second} &  \\
 & Operating point & (1, 0.75), (0.5, 1), (1, 0.5) &  \\
 & Pursuer & Direct, PD, LQR, HCW intercept, MPPI-Coast, MPPI-Flee &  \\
 & Evader & Coast, random, flee, MPPI-LQR &  \\
\addlinespace
PE-8: search & MCTS pursuer & Against coast, random, flee, MPPI-Direct, MPPI-LQR & \multirow[t]{4}{=}{$d_0=\qty{300}{\meter}$, $T=\qty{1800}{\second}$, $\Delta v_B=\qty{5}{\meter\per\second}$, $r_c=\qty{10}{\meter}$, no velocity-matching condition} \\
 & MCTS evader & Against LQR, HCW intercept, MPPI-Coast, MPPI-Flee, MCTS &  \\
 & $\rho_u$, $\rho_{\Delta v}$ & $\{0.5, 0.75, 1, 1.25, 2\}\times\{0.5, 0.75, 0.9, 1, \infty\}$ &  \\
 & HCW lookahead $T_L$ & \qty{300}{\second}, \qty{600}{\second} against the six PE evaders on the full capability sweep &  \\
\bottomrule
\end{tabular}
\end{table*}

\begin{table*}[!t]
\centering\small
\sisetup{list-final-separator={, }, list-pair-separator={, }}
\caption{Lady-bandit-guard experiments, with their identifiers. Conventions follow \Cref{tab:experiments}, with $\bar u_G=\qty{0.1}{\meter\per\second}$.}
\label{tab:experiments-lbg}
\begin{tabular}{@{}>{\raggedright\arraybackslash}p{0.15\textwidth}>{\raggedright\arraybackslash}p{0.16\textwidth}>{\raggedright\arraybackslash}p{0.39\textwidth}>{\raggedright\arraybackslash}p{0.228\textwidth}@{}}
\toprule
\textbf{Experiment} & \textbf{Varied parameter} & \textbf{Values} & \textbf{Held constant} \\
\midrule
LBG-1: capability & $\rho_u$, $\rho_{\Delta v}$ & $\rho_u\in\{0.5, 0.67, 0.8, 0.91, 1, 1.11, 1.33, 2, 4\}$; $\rho_{\Delta v}$ the same values and 0, an unlimited bandit budget & \multirow[t]{5}{=}{Guard \qty{200}{\meter} and bandit \qty{1000}{\meter} from the lady; $T=\qty{1800}{\second}$, $\Delta v_G=\qty{5}{\meter\per\second}$, $r_c=\qty{10}{\meter}$, $r_b=\qty{50}{\meter}$, no velocity-matching condition} \\
 & Guard & Coast, direct, PD, LQR, HCW intercept, MPPI-Coast, MPPI-Approach &  \\
 & Bandit & Coast, direct, LQR, HCW intercept, evade, MPPI-Direct, MPPI-LQR &  \\
\addlinespace
LBG-2: geometry & Bandit starting distance & \qtylist{0.5;1;2;4}{\kilo\meter} & \multirow[t]{5}{=}{$\rho_u=\rho_{\Delta v}=1$; guard \qty{200}{\meter} from the lady; $\Delta v_G=\qty{5}{\meter\per\second}$, $r_c=\qty{10}{\meter}$, $r_b=\qty{50}{\meter}$, no velocity-matching condition} \\
 & Horizon $T$ & \qtylist{600;1200;1800;3600}{\second} &  \\
 & Guard & Coast, direct, PD, LQR, HCW intercept, MPPI-Coast, MPPI-Approach &  \\
 & Bandit & Coast, direct, LQR, HCW intercept, evade, MPPI-Direct, MPPI-LQR &  \\
\addlinespace
LBG-3: event definition & Capture radius $r_c$ & \qtylist{5;10;20}{\meter} & \multirow[t]{5}{=}{$\rho_u=\rho_{\Delta v}=1$; guard \qty{200}{\meter} and bandit \qty{1000}{\meter} from the lady; $T=\qty{1800}{\second}$, $\Delta v_G=\qty{5}{\meter\per\second}$} \\
 & Breach radius $r_b$ & \qtylist{20;50;100}{\meter} &  \\
 & Velocity matching $v_e$ & None, \qty{0.5}{\meter\per\second} on both events &  \\
 & Guard & Coast, direct, PD, LQR, HCW intercept, MPPI-Coast, MPPI-Approach &  \\
 & Bandit & Coast, direct, LQR, HCW intercept, evade, MPPI-Direct, MPPI-LQR &  \\
\bottomrule
\end{tabular}
\end{table*}

\section{Results}\label{sec:results}
This section presents the results of the experiments in \Cref{sec:experiments} and analyzes them to understand where the choice of policy changes the outcome. For the pursuer in PE and the guard in LBG we compare two groups of policies: the feedback-control policies (direct, PD, LQR, and HCW intercept with $T_L=\qty{100}{\second}$) and the MPPI policies. In each configuration we take from each group the one policy with the highest win fraction, so the policy that best represents a group can differ from one scenario configuration to the next. The difference between the two best policies, the \emph{policy gain}, measures how much the probability of winning changes with policy choice when the rest of the scenario configuration is held fixed. The two groups are empirical envelopes over the policies we implemented, not bounds on what feedback control or MPPI can achieve. %\Cref{sec:planner-results} compares MCTS and the other HCW lookaheads separately.

\Cref{fig:capability,fig:lbg-capability,fig:geometry} map the best policy of each group and the policy gain over finer grids of configurations. Each figure names the experiment it shows by its identifier in \Cref{tab:experiments,tab:experiments-lbg}. The maps evaluate that experiment on a finer grid of the same parameters for visualization, with the same initial conditions, and give identical results in every configuration the two grids share; the counts quoted in the text come from the grids in the tables. They shade outcomes that favor the bandit, the PE pursuer or the LBG attacker, in red and outcomes that favor the guard, the PE evader or the LBG defender, in blue. Because both ratios divide the guard's capability by the bandit's, larger ratios favor the guard and the blue regions in both games. The capability maps omit the unlimited budgets, which the text reports.

\subsection{Maneuverability Decides Most Pursuit-Evasion Encounters}
\begin{figure*}[t]
\centering
\includegraphics[width=\textwidth]{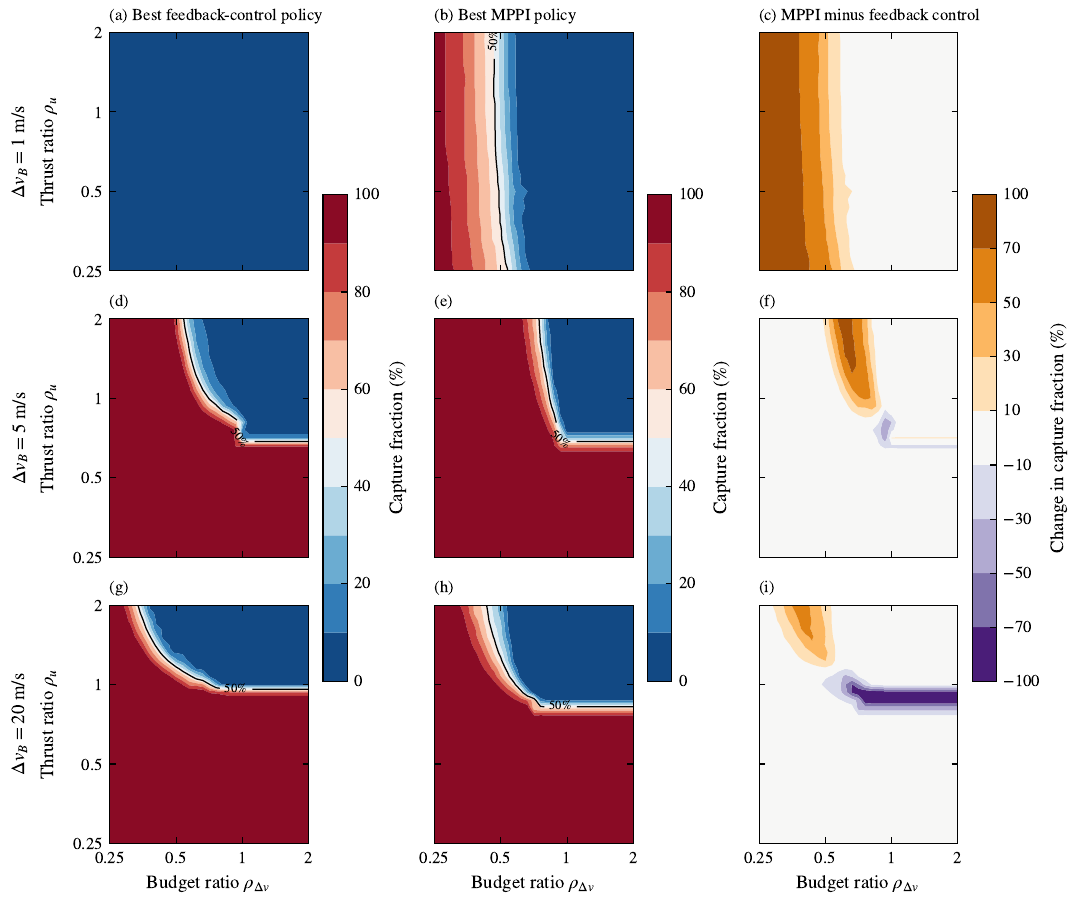}
\caption{Experiments PE-1 and PE-2. Capture fraction against an evader that flees at its maximum-thrust for the best feedback-control policy (left), the best MPPI policy (middle), and their difference, the policy gain (right), over the thrust ratio $\rho_u$ and budget ratio $\rho_{\Delta v}$. Rows show pursuer budgets of \qtylist{1;5;20}{\meter\per\second}. Red shading marks capture (a bandit win) and blue evasion (a guard win); the labeled black line is the 50\% contour. In the right column, orange marks configurations in which the best MPPI policy captures more often than the best feedback-control policy and purple those in which it captures less often. Points are evaluated on a finer grid for visualization, 31 values of each ratio with 128 encounters each, from $d_0=\qty{300}{\meter}$ with $T=\qty{1800}{\second}$.}
\label{fig:capability}
\end{figure*}

Against an evader that flees at full thrust, the ratio of propellant budgets decides the outcome over most of the capability sweep. \Cref{fig:capability}(d--f) shows the capture fraction with a \qty{5}{\meter\per\second} pursuer delta-v budget. Only 9 of the 90 configurations show a policy gain of 10\% or more in magnitude when moving from feedback-control to MPPI planning, and six of them lie at a budget ratio of 0.75. The largest gain is 64\%, at $\rho_u=1.1$ and $\rho_{\Delta v}=0.75$, where MPPI-Flee captures 77\% of encounters and the best feedback policy 13\%. The MPPI policies also capture less often than feedback control in some configurations: at $\rho_u=0.75$ and $\rho_{\Delta v}=0.9$, HCW intercept captures every evader and MPPI-Flee 68\%. A second, independent set of 128 initial conditions reproduces the result: the capture fractions of the two sets differ by 0.5\% on average over all 3,780 capability configurations, and 8 configurations show a gain of 10\% or more in magnitude, with the largest gain and loss at the same ratios.

\citeauthor{wong1967}'s work on constant-gravity games can help explain this result~\cite{wong1967}. In their work both vehicles thrust along a fixed direction, and in free space a pursuer and an evader that thrust along the line of sight gain speed in proportion to the velocity change each spends: after both budgets are spent the pursuer closes only if it has spent more, and a higher thrust limit only shortens the time to reach the final speed.  The absolute size of the budgets determines whether propellant or thrust decides. With unlimited budgets for both sides, the game becomes a race of thrust limits: the best pursuer captures every fleeing evader when $\rho_u<1$ and none when $\rho_u\ge1$. With finite budgets, three regimes follow (\Cref{fig:capability}). With a large budget of \qty{20}{\meter\per\second} the pursuer cannot spend its propellant within the horizon, so thrust decides and the boundary lies near a thrust ratio of one. With an intermediate budget of \qty{5}{\meter\per\second} propellant decides and the boundary follows the budget ratio. With a small budget of \qty{1}{\meter\per\second} the feedback-control policies capture almost no evader anywhere on the plane, and only the MPPI policies, which anticipate the budget they will have left, capture any: 95\% to 98\% of fleeing evaders at $\rho_{\Delta v}=0.25$. A larger budget is not always better for the pursuer, because at a fixed budget ratio it also lets the evader flee for longer within the fixed horizon, so a thrust advantage is needed to close the gap before the end of the game. At equal thrust limits and $\rho_{\Delta v}=0.75$, the best pursuer captures 1\%, 84\%, and 27\% of fleeing evaders with budgets of \qtylist{1;5;20}{\meter\per\second}.

This account assumes that closing a distance costs propellant in proportion to it, which holds for a pursuer that thrusts along the line of sight and not for one that uses the orbital dynamics. Two coasting vehicles that start \qty{300}{\meter} apart in the cross-track direction come within the capture radius after \qty{1430}{\second} with no maneuver at all. HCW intercept reaches a coasting evader for less propellant as its lookahead grows, spending \qtylist{3.2;1.8;1.0}{\meter\per\second} on average with lookaheads of \qtylist{100;300;600}{\second}. Within the \qty{1800}{\second} horizon these frugal transfers lose to faster ones, since the \qty{600}{\second} lookahead captures 3\% of coasting evaders where the shorter lookaheads capture all of them. A longer or repeated game, in which propellant spent early is unavailable later, would favor them, which we did not test for brevity as these games would warrant a complete study of their own. Against coasting and random evaders the propellant constraint is rarely the limiting factor at \qty{5}{\meter\per\second}, where the best pursuer captures every coasting evader and at least 79\% of random evaders in every configuration.

\subsection{Policy Choice Decides Lady-Bandit-Guard}\label{sec:lbg-results}
\begin{figure*}[t]
\centering
\includegraphics[width=\textwidth]{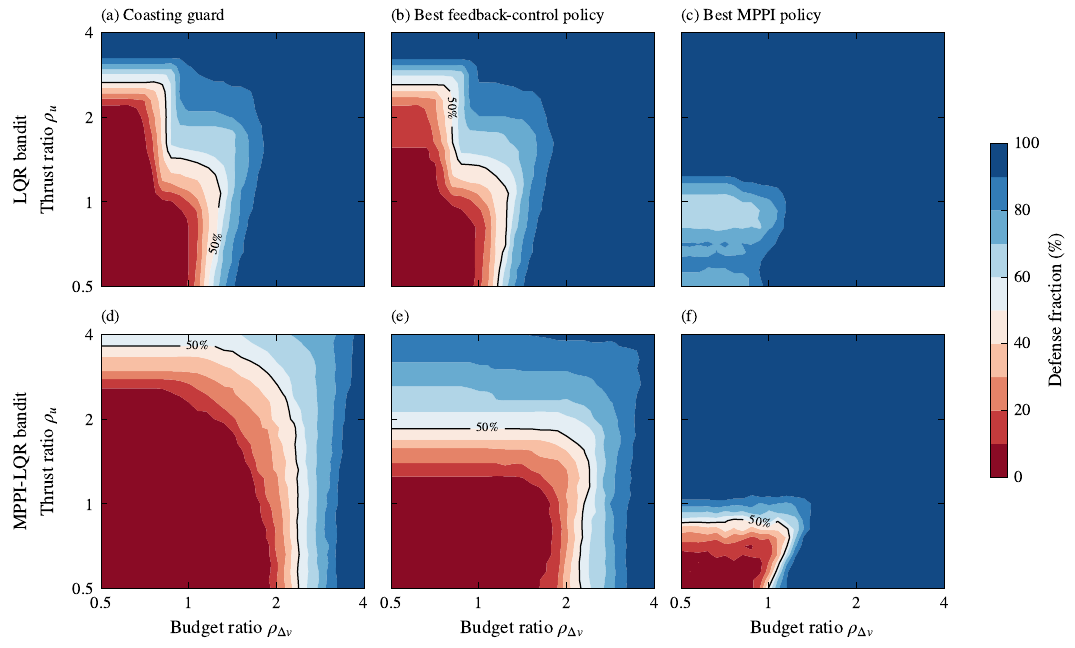}
\caption{Experiment LBG-1. Defense fraction in LBG of a coasting guard (left), the best feedback-control guard (middle), and the best MPPI guard (right) against the LQR bandit (top) and the MPPI-LQR bandit (bottom), over the thrust ratio $\rho_u$ and budget ratio $\rho_{\Delta v}$, guard over bandit. Blue shading marks defense (a guard win) and red breach (a bandit win), so the guard's advantage grows toward larger ratios; the labeled black line is the 50\% contour. The guard budget is \qty{5}{\meter\per\second}; the bandit starts \qty{1}{\kilo\meter} and the guard \qty{200}{\meter} from the lady. Points are evaluated on a finer grid for visualization, 31 values of each ratio with 128 encounters each.}
\label{fig:lbg-capability}
\end{figure*}

In LBG the same capability sweep gives the opposite answer: against the bandits that threaten the lady, the choice of policy decides the outcome over most configurations. \Cref{fig:lbg-capability} shows the defense fraction of an idle, coasting guard, of the best feedback-control guard, and of the best MPPI guard against the LQR and MPPI-LQR bandits. At equal thrust limits and delta-v budgets against the MPPI-LQR bandit, they defend 0\%, 1\%, and 83\% of encounters. Against the four bandits that breach most often (LQR, HCW intercept, MPPI-Direct, and MPPI-LQR), the best MPPI guard defends 10\% or more above the best feedback-control guard in 56 to 74 of the 90 configurations and falls that far below it in none. The direct and evasive bandits rarely breach even an idle guard, so the guard's policy matters little against them.

The feedback-control guards do not fail for lack of propellant. Against the LQR bandit at equal thrust limits and delta-v budgets, the LQR and HCW intercept guards pass the bandit \qtyrange{35}{40}{\meter} away, outside the \qty{10}{\meter} capture radius, and the bandit breaches in 77\% to 79\% of encounters. Giving these guards an unlimited delta-v budget leaves their defense fraction unchanged. The MPPI-Approach guard, which forecasts the bandit's approach to the lady, spends less than they do, \qty{3.5}{\meter\per\second}, and closes to \qty{9}{\meter}. Changes in outcomes due to varying the budget ratio in \Cref{fig:lbg-capability} therefore come through effects on the bandit capabilities: a feedback-control bandit with at most half the guard's delta-v budget, or one that starts \qty{2}{\kilo\meter} or more from the lady, rarely breaches even an idle guard. The MPPI bandits breach in both cases, in up to 91\% of encounters with half the guard's budget and in 60\% from \qty{2}{\kilo\meter}. A horizon longer than \qty{1200}{\second} changes no defense fraction, because by then every bandit has breached, been captured, or spent its budget.

\subsection{Distance and Time Set the Reach of Each Policy}
\begin{figure*}[t]
\centering
\includegraphics[width=\textwidth]{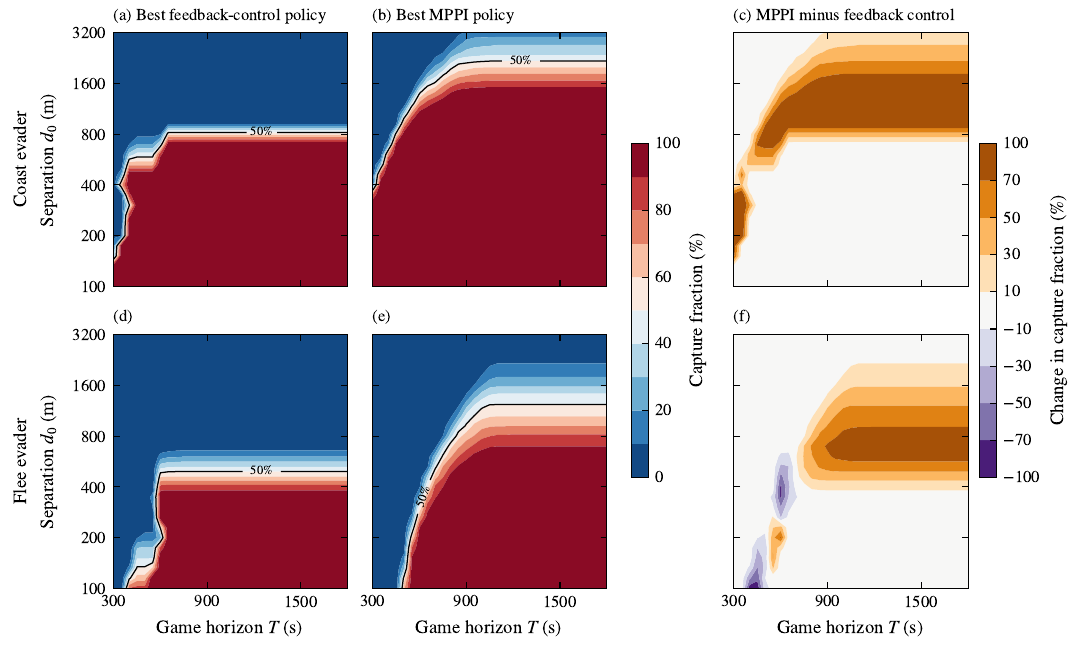}
\caption{Experiment PE-3. Capture fraction over initial separation and game horizon for the best feedback-control policy (left), the best MPPI policy (middle), and their difference (right). The top row uses a coasting evader with $\rho_u=1$ and $\rho_{\Delta v}=0.75$; the bottom row a fleeing evader with $\rho_u=1$ and $\rho_{\Delta v}=0.5$. Colors and the 50\% contour follow \Cref{fig:capability}. The pursuer budget is \qty{5}{\meter\per\second}. Points are evaluated on a finer grid for visualization, 26 separations by 31 horizons with 128 encounters each. Horizons of \qty{3600}{\second} and \qty{5400}{\second} give the same values as \qty{1800}{\second}.}
\label{fig:geometry}
\end{figure*}

The initial separation and the game horizon determine which pursuers can reach the evader at all, and the MPPI policies reach farther (\Cref{fig:geometry}). Against a coasting evader, with a \qty{5}{\meter\per\second} pursuer budget and a \qty{1800}{\second} horizon, the best feedback-control pursuer captures about half of the encounters from \qty{800}{\meter} and none from \qty{1.6}{\kilo\meter}, while the best MPPI pursuer captures every encounter from \qty{800}{\meter} and most from \qty{1.6}{\kilo\meter}. This is consistent with the propellant mechanism described above, since a pursuer that must travel farther needs more of its budget to arrive. The MPPI pursuers reach farther by spending that budget more efficiently, not by taking slower transfers: from \qty{800}{\meter} MPPI-Flee captures every encounter at a median of \qty{450}{\second} and spends \qty{4.0}{\meter\per\second} on average, where the LQR pursuer spends nearly its whole budget and captures 57\% at a median of \qty{620}{\second}.

In experiment PE-3, a horizon longer than \qty{1800}{\second} changes the result of neither group. A horizon that is too short can reverse the ranking of policies. From \qty{100}{\meter} against a fleeing evader, direct pursuit captures 62\% of encounters within \qty{600}{\second}, while HCW intercept, LQR, and MPPI-Flee capture at most 24\%; within \qty{900}{\second} those three capture nearly every encounter and direct pursuit still 62\%. Their captures begin at about \qty{590}{\second}, so a \qty{600}{\second} horizon ends before most of them and ranks the crudest policy first.

While these results are likely unsurprising for anyone experienced with astrodynamics and spacecraft control, it is an important consideration when evaluating policies in orbital game scenarios that the selection of horizon and initial separation for a constant set of spacecraft capabilities can significantly affect, and even entirely determine, the win-rate.

\subsection{Information Changes Which Policy Wins}
\begin{figure*}[t]
\centering
\includegraphics[width=\textwidth]{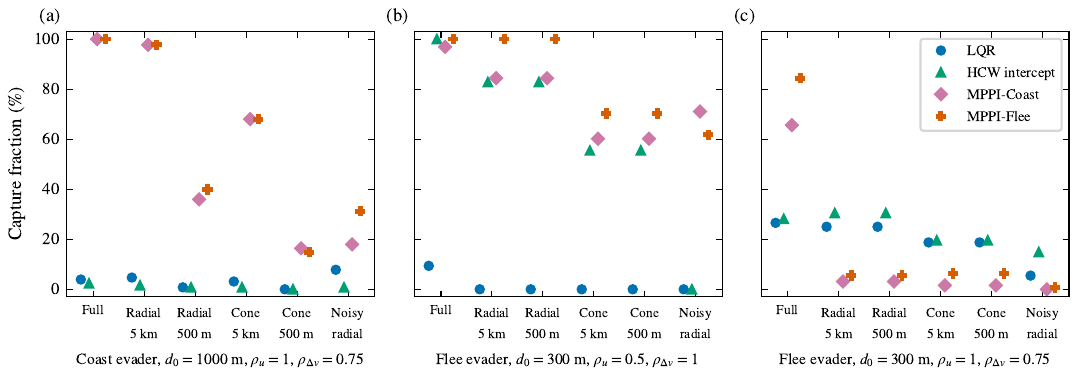}
\caption{Experiment PE-4. Capture fraction by sensor for four pursuers. Radial and conical (\ang{10} half-angle) sensors have the stated range and a noise of 1\% of range; the noisy radial sensor has a range of \qty{5}{\kilo\meter} and 10\% noise. With a sensor, both sides sense, and each encounter starts from a Gaussian prior error with standard deviations of $0.1d_0$ in position and \qty{0.01}{\meter\per\second} in velocity per axis. Each point averages the 128 encounters of one configuration.}
\label{fig:information}
\end{figure*}

Including imperfect information and sensor noise lowers the capture fraction by different amounts for different policies as seen in \Cref{fig:information}. The capture fraction of an MPPI policy follows the quality of its sensor. Against a coasting evader \qty{1000}{\meter} away, MPPI-Flee captures every encounter with full information, 68\% with a \ang{10} cone sensor, and 40\% with a radial sensor of \qty{500}{\meter} range, while the feedback-control pursuers capture few encounters under any sensor.

Which group is more robust depends on capability. With a thrust advantage ($\rho_u=0.5$, \Cref{fig:information}(b)), the MPPI policies degrade less than HCW intercept. At the \emph{contested operating point} ($\rho_u=1$, $\rho_{\Delta v}=0.75$), where neither side wins by capability alone and the capture fraction against the fleeing evader ranges from 0\% to 84\% with the pursuer's policy, the ranking reverses (\Cref{fig:information}(c)): MPPI-Flee falls from 84\% to 5\% with a radial sensor whose estimate is in error by \qty{3.5}{\meter} on average, while HCW intercept is unchanged near 30\%. Changing which side senses, the prior error, and the measurement noise one at a time shows that the error of the pursuer's own estimate is the main cause.

The full-information baseline of the other experiments is therefore optimistic. Repeating the PE capability experiment with a \ang{10} cone sensor on both sides lowers capture over most of the sweep; the best pursuer captures 77\% of coasting evaders where with full information it captured all of them. The choice of policy still changes the capture fraction by 10\% or more in at most 15 of the 90 configurations against five of the six evaders, but the group that wins those configurations changes: against the fleeing evader the feedback-control group is better in 9 of the 15, where with full information the MPPI group was better in 7 of 9.

\subsection{Policy Settings Interact with the Scenario}\label{sec:planner-results}
\begin{figure*}[t]
\centering
\includegraphics[width=\textwidth]{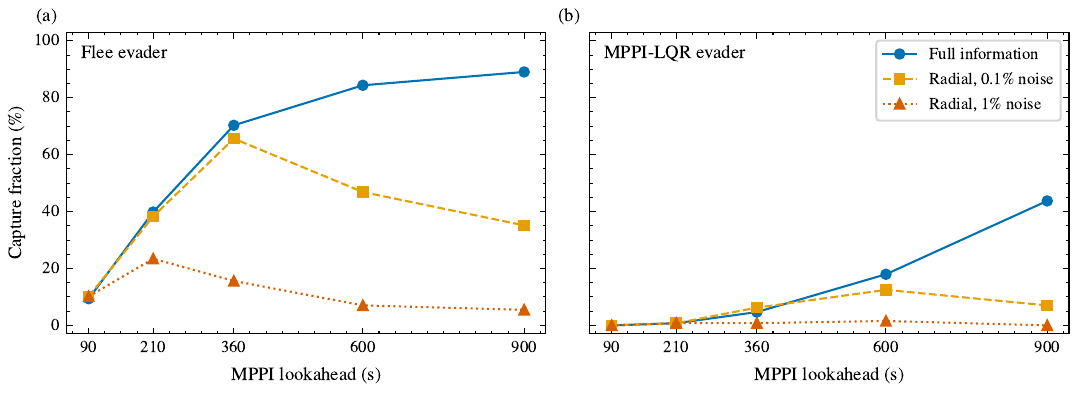}
\caption{Experiment PE-5. Capture fraction of MPPI-Flee by planning lookahead at the contested operating point ($\rho_u=1$, $\rho_{\Delta v}=0.75$, $d_0=\qty{300}{\meter}$) for three levels of pursuer information. Only the pursuer senses, and the MPPI-LQR evader keeps a \qty{600}{\second} lookahead; each point averages the 128 encounters of one configuration.}
\label{fig:lookahead}
\end{figure*}

The settings of a policy are part of the scenario as much as the vehicle capabilities are. \Cref{fig:lookahead} shows the capture fraction of MPPI-Flee at the contested operating point as its lookahead varies. With full information a longer lookahead helps against the fleeing evader, raising capture from 9\% at \qty{90}{\second} to 89\% at \qty{900}{\second}. With a radial sensor the best lookahead shortens as the noise grows, to \qty{360}{\second} with 0.1\% noise and \qty{210}{\second} with 1\%. HCW intercept is as sensitive to its lookahead: with \qty{600}{\second} it captures 3\% of coasting evaders within the horizon, and with 100 or \qty{300}{\second} all of them.

An MPPI policy's forecast of its opponent changed its outcome more than its sampling effort did. An evader that forecasts direct pursuit escapes LQR and HCW intercept but is captured by the MPPI pursuers in most encounters when it has less thrust. An evader that forecasts LQR pursuit escapes the MPPI pursuers when it has at least three quarters of their thrust but is captured by HCW intercept when it has half. Neither forecast is best against every pursuer, and raising the evader's sample count from 64 to 1,024 changes its capture fraction by at most 15\%. A second search-based policy, MCTS, gives results between those of the feedback-control policies and MPPI (\Cref{tab:mcts}).

\begin{table}[!htb]
\centering\small
\caption{Experiment PE-8. Capture fraction, in percent, of three pursuers against each evader, averaged over 25 configurations of the PE capability sweep ($\rho_u\in\{0.5,0.75,1,1.25,2\}$, $\rho_{\Delta v}\in\{0.5,0.75,0.9,1,\infty\}$).}
\label{tab:mcts}
\begin{tabular}{@{}lrrr@{}}
\toprule
\textbf{Evader} & \textbf{HCW intercept} & \textbf{MPPI-Flee} & \textbf{MCTS} \\
\midrule
Coast & 100 & 100 & 100 \\
Random & 98 & 98 & 96 \\
Flee & 46 & 51 & 45 \\
MPPI-Direct & 1 & 62 & 31 \\
MPPI-LQR & 27 & 42 & 32 \\
MCTS & 21 & 52 & 34 \\
\bottomrule
\end{tabular}
\end{table}

\subsection{The Capture Definition Changes the Winner}\label{sec:capture-results}
Unsurprisingly, how a study defines capture changes which policy wins; however, different policies are more or less robust to changes in the definition. At the contested operating point against the fleeing evader, the best feedback pursuer captures 13\% of encounters with $r_c=\qty{5}{\meter}$ and 84\% with $r_c=\qty{50}{\meter}$, while the best MPPI policy captures 84\% to 88\% at every radius, so the policy gain falls from 70\% to 3\%. Requiring matched velocity makes the same game one-sided: with a velocity-matching condition of $v_e=\qty{0.5}{\meter\per\second}$ no pursuer captures more than 9\% of fleeing evaders, since a pursuer with equal thrust must match the velocity of an evader that has been accelerating away as well as close on it. In LBG the same condition on both events lowers the largest advantage of the best MPPI guard over the best feedback-control guard, across all bandits and radii, from 91\% to 28\%. How often the game tests for capture matters as well: testing only at the end of each \qty{10}{\second} step changes the feedback policies' capture fractions by at most 8\% but costs the MPPI policies, whose trajectories pass through the evader at speed, up to 69\%.

\section{Conclusions}\label{sec:conclusions}

This paper introduced OrbitalGym, a JAX environment for orbital games, and used it to study where the choice of policy changes the outcome of two close-range games. In pursuit-evasion the ratio of propellant budgets organized the outcome against the fleeing evader, and against five of six evaders the choice between the tested MPPI and feedback-control policies mattered little: with a \qty{5}{\meter\per\second} pursuer budget, it changed the capture fraction by 10\% or more in at most 15 of 90 capability configurations. In lady-bandit-guard the choice of policy mattered far more: the best MPPI guard raised the guard's defense by 10\% or more in 56 to 74 of 90 configurations against each of the four bandits that breached most often. The reasons differ between the games. In pursuit-evasion the choice of policy mattered where propellant bound the pursuer, which the MPPI policies anticipate over their lookahead and the feedback-control policies do not. In lady-bandit-guard the feedback-control guards missed an accelerating bandit even with unlimited propellant, while the MPPI guard that forecasts the bandit's approach intercepted it.

The remaining experiments each carry a lesson for evaluating policies in orbital games. The initial separation and the horizon determine which policies can reach their opponent at all, and a horizon that is too short can reverse their ranking. Full information is an optimistic baseline: sensing lowered the capture fraction and changed which group of policies won. A policy's lookahead and its forecast of the opponent changed the outcome more than its sampling effort did. The capture radius and a velocity-matching condition changed which policy won. A comparison of decision methods for orbital games therefore needs to report these parameters with the result and to test whether the result holds when they change.

Future work could extend the study in four directions. Far-range encounters call for nonlinear dynamics and could draw on the reachable-set analysis of~\cite{venigalla2021}. Games with more than one vehicle per side raise the question of when communication between teammates changes the outcome, which a semi-decentralized formulation makes explicit. Improving the realism of the simulation through the addition of power, battery, and eclipse models would add a resource that competes with sensing and thrust. Finally, learned policies could be placed in the same maps to test whether their gains persist outside the configurations in which they are trained.

\appendix{ A: Policy Formulations}
\refstepcounter{section}\label[appendix]{app:controllers}
This appendix gives the full formulation of each policy of \Cref{sec:methods}. Each command is limited by \Cref{eq:limits} before it is applied, and a policy that plans on a belief uses the estimate $\hat x_k$ in place of $x_k$. The pursuit laws are written for the PE pursuer, the bandit, with $x=s_B-s_G$ the pursuer's state relative to the evader, $r$ its position components, and $v$ its velocity components. The LBG guard applies the same laws to its state relative to the bandit, $-x$, with its own thrust limit $\bar u_G$, and the LBG bandit applies them to its state relative to the lady.

\subsection{Non-Reactive and Feedback Policies}\label{app:feedback}
The random policy draws a direction uniformly on the unit sphere and a magnitude $\bar u\,U^{1/3}$ with $U\sim\mathcal U(0,1)$, which is uniform in the ball of admissible impulses, independently at every step. The direct pursuer commands $u_B=-\bar u_B\,r/\|r\|$, and the fleeing evader commands $u_G=-\bar u_G\,r/\|r\|$, which increases the separation because $r$ points from the evader to the pursuer. The transverse evader applies full thrust along the unit vector $-\hat r\times e_N$, where $\hat r=r/\|r\|$, with the radial axis replacing $e_N$ when the line of sight is within \ang{25.8} of the cross-track direction. The evading bandit commands the direct impulse toward the lady, rotated away from the guard's line of sight whenever the guard is within \qty{300}{\meter}.

The PD pursuer applies proportional and derivative feedback on the relative position $r$ and the relative velocity $v$,
\begin{equation}
u_B=-\Delta t\left(K_pr+K_dv\right)
\label{eq:pd}
\end{equation}
with $K_p=\qty[per-mode=power]{e-4}{\per\second\squared}$ and $K_d=\qty[per-mode=power]{0.02}{\per\second}$, which, for the continuous-time double integrator in free space without saturation, give a damping ratio of one and a time constant of \qty{100}{\second}.

The LQR pursuer uses the gain
\begin{equation}
K=\left(Q_u+\Gamma^\top P\Gamma\right)^{-1}\Gamma^\top PA
\label{eq:lqr}
\end{equation}
where $A=A(\Delta t)$, $\Gamma=A(\Delta t)B$, and $P$ solves the discrete algebraic Riccati equation for the state and input weights
\begin{equation}
Q_x=\begin{bmatrix}10^{-4}I_3&0\\0&I_3\end{bmatrix},\qquad Q_u=100\,I_3
\label{eq:lqr-weights}
\end{equation}
in SI units, which scale the position by \qty{100}{\meter}, the velocity by \qty{1}{\meter\per\second}, and the impulse by \qty{0.1}{\meter\per\second}. The regulator assumes a coasting evader. Under a constant evader thrust it has a steady-state offset, because it has no integral action.

The HCW intercept pursuer partitions the transition matrix over the lookahead $T_L$ into position and velocity blocks, $A(T_L)=\bigl[\begin{smallmatrix}\Phi_{rr}&\Phi_{rv}\\\Phi_{vr}&\Phi_{vv}\end{smallmatrix}\bigr]$. The impulse that brings the relative position to zero at $T_L$ against a coasting evader is
\begin{equation}
u_B=-\Phi_{rv}^{-1}\left(\Phi_{rr}r+\Phi_{rv}v\right)
\label{eq:intercept}
\end{equation}
The pursuer recomputes \Cref{eq:intercept} at every step with the same $T_L$, so the intended arrival time recedes with the game rather than being held fixed. Against a coasting target, while the command is not saturated and propellant remains, each impulse sets the closing speed to roughly $\|r\|/T_L$, so the separation decays approximately as $e^{-t/T_L}$ and reaches the capture radius after about $T_L\ln(d_0/r_c)$. Holding the arrival time fixed instead would aim to arrive at $T_L$ exactly, and would need large corrections for any error that remains as the arrival time approaches. The command imposes no terminal velocity and is not fuel-optimal.

\subsection{Model Predictive Path Integral Planning}\label{app:mppi}
\Cref{alg:mppi} lists one MPPI decision for side $i$ against opponent $j$. Each sample perturbs the nominal sequence $\mu_{1:H}$ with Gaussian noise drawn independently for each of the $H/m$ segments and held for the $m$ steps of the segment, and \textsc{Clip} scales an impulse down to the thrust limit along its direction. The update in line 11 is written for a side whose own state enters the relative state with a positive sign, the bandit for $x=s_B-s_G$; the other side's update subtracts its own impulse instead. The stage costs $c_i=-R_i$ are those of \Cref{eq:rewards}. We standardize the costs to $\tilde J^n$ before weighting, and budgets limit the sampled sequences and the forecast opponent as in the game. By default, all MPPI policies use $N=256$ samples, a horizon of $H=60$ steps (\qty{600}{\second}) in which each noise perturbation is held for $m=3$ steps, noise scale $\beta=0.5$, temperature $\lambda=0.1$, discount $\gamma=0.99$ per step, and event bonus $b=10$. We did not tune these values per configuration.

\begin{algorithm}[!htb]
\caption{MPPI step for side $i$ against opponent $j$}
\label{alg:mppi}
\begin{algorithmic}[1]
\Function{Plan}{$\hat x$, $\mu_{1:H}$, $\bar u_i$, $\bar u_j$}
\For{$n=1,\dots,N$}
  \For{$q=1,\dots,H/m$}
    \State $\epsilon^n_q\sim\mathcal N\left(0,\beta^2\bar u_i^2I_3\right)$ \Comment{one draw per segment}
    \State $\epsilon^n_h\gets\epsilon^n_q$ for $h=(q-1)m+1,\dots,qm$
  \EndFor
  \State $u^n_h\gets\textsc{Clip}(\mu_h+\epsilon^n_h,\bar u_i)$ for $h=1,\dots,H$
  \State $x\gets\hat x$, $J^n\gets0$
  \For{$h=1,\dots,H$}
    \State $u_j\gets\pi_j(x)$ \Comment{opponent forecast}
    \State $x\gets A(\Delta t)\left(x+B\left(u^n_h-u_j\right)\right)$
    \State $J^n\gets J^n+\gamma^{h-1}c_i(x)$
  \EndFor
\EndFor
\State $w^n\gets\exp(-\tilde J^n/\lambda)/\sum_{n'}\exp(-\tilde J^{n'}/\lambda)$
\State $\mu_{1:H}\gets\textsc{Clip}(\sum_n w^nu^n_{1:H},\bar u_i)$
\State \Return $\mu_1$, $[\mu_{2:H},\mu_H]$
\EndFunction
\end{algorithmic}
\end{algorithm}

\subsection{Monte Carlo Tree Search}\label{app:mcts}
The search runs the Stochastic MuZero policy of mctx~\cite{mctx,antonoglou2022stochastic} with 256 simulations per decision and a lookahead of \qty{600}{\second}. Each tree edge holds one of the nine impulses of \Cref{tab:policies} for $m=3$ steps and passes through a chance node with one branch per opponent policy, drawn uniformly, so the value of an action averages over the opponent's possible policies. The reward of an edge is the discounted sum of the stage rewards $R_i$ of \Cref{eq:rewards} over its three steps, with the per-step discount $\gamma=0.99$ and event bonus $b=10$ of MPPI, and an edge that ends the game has zero continuation. A new leaf is valued by one Monte Carlo rollout in which the searching side follows the rollout policy, HCW intercept with $T_L=\qty{100}{\second}$ for the pursuer and fleeing for the evader, against an opponent policy drawn from the same set, truncated where the lookahead ends. Each of the 256 simulations evaluates two rollouts of up to 60 steps, one at the action's afterstate and one at the chance outcome, so a decision simulates about 32,000 steps, roughly twice MPPI's $N\times H=15{,}360$. The side executes the most visited root action.

\bibliographystyle{IEEEtranN}
\bibliography{paper/references}

\thebiography
% Insert immediately after \thebiography in paper/manuscript.tex.
% Source and author-review notes: paper/bios/README.md.
\begingroup
% The class specifies both width and height; retain the photos' proportions.
\setkeys{Gin}{keepaspectratio}

\begin{biographywithpic}{Duncan Eddy}{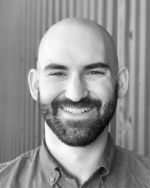}
is a staff researcher in the Stanford Intelligent Systems Laboratory (SISL). His research focuses on decision-making in safety-critical, climate, and space systems. He received a B.S. in Mechanical Engineering from Rice University in 2013 and a Ph.D. in Aerospace Engineering from Stanford University in 2021. Prior to returning to Stanford, he was the Director of Space Operations at Capella Space Corporation, where he built a fully automated constellation tasking and delivery system. He later founded and led the Constellation Management and Space Safety Organization at Amazon's Project Kuiper.
\end{biographywithpic}

% Draft adapted from Mahdi's website; Mahdi should review before submission.
\begin{biographywithpic}{Mahdi Al-Husseini}{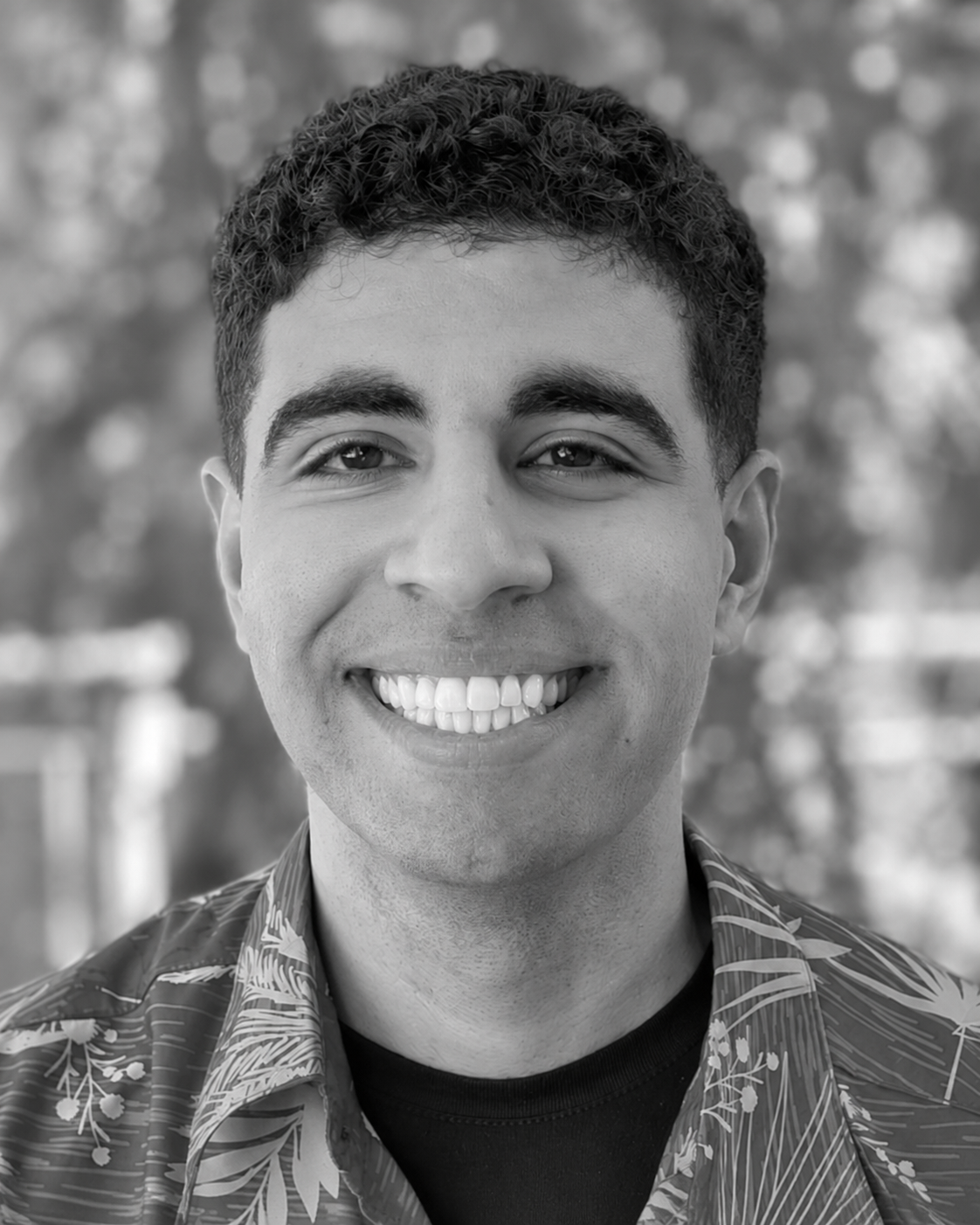}
is a Ph.D. candidate in the Stanford Intelligent Systems Laboratory (SISL). His research develops models and algorithms for semi-decentralized multi-agent systems operating under communication constraints. He received B.S. degrees in Biomedical Engineering and Public Policy and an M.S. in Computer Science from the Georgia Institute of Technology, and an M.S. in Aeronautics and Astronautics from Stanford University. He is a U.S. Army captain and an experimental test pilot in training.
\end{biographywithpic}

\begin{biographywithpic}{Mykel J. Kochenderfer}{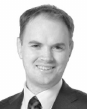}
is an Associate Professor of Aeronautics and Astronautics and Associate Professor, by courtesy, of Computer Science at Stanford University. He is the director of the Stanford Intelligent Systems Laboratory (SISL), conducting research on advanced algorithms and analytical methods for the design of robust decision-making systems. Prior to joining the faculty in 2013, he was at MIT Lincoln Laboratory, where he worked on airspace modeling and aircraft collision avoidance. He received his Ph.D. from the University of Edinburgh in 2006, where he studied at the Institute of Perception, Action and Behaviour in the School of Informatics. He received B.S. and M.S. degrees in computer science from Stanford University in 2003.
\end{biographywithpic}
\endgroup

\end{document}